\documentclass[12pt]{article}
     \usepackage{amssymb,amsmath}
     \usepackage[english]{babel}
     \usepackage{graphicx}
     
\begin{document}
\baselineskip 5mm
\thispagestyle{empty}
\title{\vspace{-2cm}
Artificial magnetism in theory of wave multiple scattering 
by random discrete non-magnetic conducting media}

\vspace{1cm}

\author{Yu.N.Barabanenkov$^1$, M.Yu.Barabanenkov$^2$, S.A.Nikitov$^1$ 
     \\     
     \\      
       \small \em $^1$V.A Kotelnikov Institute of Radioengineering and Electronics, Russian Academy of Sciences\\
        \small \em Mohovaya 11, 103907 Moscow, GSP--3,  Russia
     \\
       \small \em $^2$Institute of Microelectronics Technology, Russian Academy of Sciences\\
        \small \em 142432 Chernogolovka, Moscow Region, Russia}
\date{}
 \maketitle
 
\parbox{16cm}{\small We show that technique of Dyson equation in wave multiple scattering 
by spatially disordered discrete medium statistical theory leads directly 
to a dielectric permittivity tensor, which is characterized by spatial 
dispersion and obeys the generalized Lorentz--Lorenz formula. 
    Introduced via this spatial dispersion an effective magnetic permeability 
    demonstrates the diamagnetic property in limit of independent strongly reflected 
    non--magnetic small spherical particles in accordance with earlier intuitive predictions. 
The revealed physical nature of the effective diamagnetic property consists in that   
electric and magnetic dipoles induced in a particle by wave scattering give different 
contribution into the transverse and longitudinal components of the effective 
dielectric permittivity.  
    Besides as appeared the diamagnetism under study is enhanced by appearance 
    of additional effective dielectric displacement current in the medium.}

\section{Introduction}

Effective medium theory or homogenization approach is a powerful way to handle the radiative  
properties of composite materials, such as, e.g., cosmic dust, aerosols, porous media, 
and microstructured  materials or "metamaterials". 
      Several interesting potentials of these metamaterials have been demonstrated 
      and/or suggested, including the realization of media with negative index of refraction 
      \cite{SS01,SM00} and artificial magnetism from conductors \cite{SCHELK52,KSH94,P99}. 
Even though some fundamental concepts about the metamaterials such as negative permittivity 
and negative permeability are widely accepted and relatively well understood today, 
the  problem to define unambiguously these parameters by systematic and clear way 
is not yet totally resolved. 
     For example, the introducing artificial magnetism by \cite{SCHELK52,KSH94,P99} 
     is rather intuitive one. 
The two variable asymptotic expansion procedure \cite{BENS78}, which makes homogenization of 
Maxwell's equations in a periodically composite under condition that a period tends to zero, 
is presented as purely mathematic one and does not discuss the problem of artificial 
magnetism from non--magnetic conductors. 
     The author \cite{SILV07} has derived a productive transformed integral equation 
     for the electric field in periodically microstructured composite, with inhomogeneous term 
     being proportional to electric field averaged over unit cell of the microstructure 
     and integrating over unit cell in the integral term.  
This equation was solved numerically to get the spatially dispersive effective dielectric 
permittivity tensor and to consider, in particular, effective magnetic permeability 
via the spatial dispersion of dielectric permittivity according to \cite{LL04}.  
    In \cite{SILV07S}, a semi--intuitive approach is constructed to solve analytically 
    the transformed integral equation \cite{SILV07}, with obtaining some generalization 
    of classical Lorentz--Lorenz formula for local electric field and similar formula for 
    local magnetic field that gave possibility to obtain a corresponding generalization 
    of the Clausius--Mossotti formulas for the classical "local" effective permittivity 
    and permeability.    
     
So far the approaches were discussed to homogenizing the Maxwell's equations in a periodically 
composites, with averaging over crystal cell. 
    The aim of this paper is to show the systematic and clear way for effective magnetism
    with diamagnetic property definition in framework of electromagnetic wave multiple scattering
    theory by spatially disordered composites consisted of non--magnetic particles, each of which
    is characterized by given dielectric permittivity and conductivity.  
The wave multiple scattering theory \cite{FRISH66,FRISH67} starts with considering the wave
scattering by a single particle that can be described in terms of particle tensor electric
field scattering operator \cite{NEW66}.
   The principal point for our consideration is that this scattering operator is restored
   from Mie solution \cite{BORNW64} for electromagnetic wave scattering field in all points 
   of spherical particle outside, including near zone, in part of electric dipole and magnetic 
   dipole scattering, that is with accounting the spherical harmonics of unit angular momentum. 
As this takes place, the contribution of electric dipole scattering into electric field scattering 
operator in momentum space representation is appeared to be not dependent on momentums 
whereas the contribution of magnetic dipole scattering is a bilinear function of scattering and 
incident momentums what is very suitable for effective magnetic permeability evaluation 
via effective dielectric permittivity tensor with quadratic spatial dispersion \cite{LL04}. 
    The effective dielectric permittivity tensor is evaluated in our formalism from generalized 
    Lorentz-Lorenz formula, which is derived by homogenization of stochastic Maxwell's equations 
    for electromagnetic field in random medium under consideration with using the technique of 
    Dyson equation \cite{FRISH66,FRISH67} and averaging procedure over statistical ensemble of 
    particles.
Our observation the different contributions of a particle electric and magnetic dipole scattering 
into the transverse and longitudinal components of the effective dielectric permittivity 
is fundamental for appearance of an artificial effective diamagnetic properties in statistical 
ensemble of non--magnetic conducting particles. 
    The authors \cite{AG06} discuss similar phenomena in contribution of the electric--dipole and 
    the magnetic--dipole transitions to the dielectric tensor in the framework of molecular light 
    scattering.

In what follows we present the materials of our paper making accent on technique of generalized   
Lorentz--Lorenz formula related to statistical ensemble of non--magnetic conductive particles 
that can be especially useful at appearance a method \cite{PATENT09} of artificial dielectric 
disordered material manufacturing with controlled parameters.

\section{Homogenization of stochastic Maxwell's equations}  
We start with stochastic Maxwell's equations for monochromatic electromagnetic wave field 
\begin{equation}\label{STM}
\text{rot} \bf{E} = i \, \frac {\omega}{c} \, \bf{B},
\hskip 2em
\text{rot} \bf{B} = - i \, \frac {\omega}{c} \, \hat\varepsilon \, \bf{E}
\end{equation}
in homogeneous background (host) medium with frequency $\omega$, 
constant dielectric permittivity $\varepsilon_0$ and unit magnetic permeability $\mu$ = 1, 
where there are placed randomly particles with given dielectric permittivity  $\varepsilon_1$, 
specific conductivity $\sigma_1$ and unit magnetic permeability.
     Symbol  $\hat\varepsilon (\bf {r})$ denotes stochastic complex dielectric permittivity 
     of the medium under study, being equal to complex dielectric permittivity 
     $\hat\varepsilon_1$ = $\varepsilon_1$ + $i4\pi\sigma_1 / \omega$  inside a single particle 
     and $\varepsilon_0$ outside the particles. 
The problem of the random medium effective parameters is resolved by averaging the Eqs.(\ref{STM})
over particle spatial distribution statistical ensemble, with the averaging denoted 
as $\left\langle \dots \right\rangle$.
     In particular, an effective dielectric permittivity of the medium $\varepsilon_{eff}$ 
     is obtained according to \cite{FRISH66,FRISH67,FINK68,BAR76}  via a "decoupling" 
\begin{equation}\label{DCPL}
\left\langle \hat\varepsilon (\bf{r}) E_\alpha (\bf{r}) \right\rangle 
= 
\int \, d\bf{r}' \, \varepsilon^{eff}_{\alpha\beta} (\bf{r},\bf{r}') \, 
\left\langle E_\beta (\bf{r}') \right\rangle 
\end{equation}
the averaged product $\left\langle \hat\varepsilon E_\alpha \right\rangle$ of the random medium complex 
dielectric permittivity and the random electric field vector $E_\alpha (\bf{r})$
with components $\alpha$ = 1, 2, 3. 
       As one can see, effective dielectric permittivity of the random medium is a tensor operator, 
       with tensor indices being denoted by Greek subscripts and implied agreement about summing 
       over repeated subscripts. 
The decoupling (\ref{DCPL}) is realized with the aid of the Dyson equation \cite{FRISH66,FRISH67}, 
which is obtained via ensemble averaging the integral equation for stochastic electric field
and written in a symbolic operator form as
\begin{equation}\label{STEF}
\left\langle E \right\rangle = E^{(0)} \, + G^{(0)} M  \left\langle E \right\rangle 
\end{equation}

      In the right hand side of this equation $E^{(0)}_\alpha (\bf{r})$ denotes  the  incident 
      electromagnetic wave electric field;  
      the tensor Green function of electric field in the host medium is given by 
      $G^{(0)}_{\alpha\beta}(\bf{r})$ = $\left ( \delta_{\alpha\beta} \right.$ 
      + 
      $\left.\bigtriangledown_\alpha \bigtriangledown_\beta /k^2_0 \right )$ $G_0(r)$,
      with
      $k_0$ = $\varepsilon^{1/2}_0$ $\omega /c$ 
      being the wave number in the host medium and 
      $G_0(r)$ = $\text{exp} \left ( i k_0 r  \right )$/$\left ( -4\pi r  \right )$
      a scalar Green function; 
      $M_{\alpha\beta} (\bf{r},\bf{r}')$ is the tensor mass operator.
The effective dielectric permittivity and mass operator are connected between them 
by operator relation,
$\varepsilon_{eff}/\varepsilon_0$ = $I$ - $M/k^2_0$, 
where $I$ is  the identical operator.

\section{Lorentz--Lorenz formula}
The tensor Green function $G^{(0)}$ of electric field in the host homogeneous medium 
is strongly singular in the origin point and decomposed \cite{FINK68,FINK64,MALET05} 
into delta Dirac function term $1/(3k^2_0)I$ and the principal part $\tilde G^{(0)}$. 
    Substituting this singular tensor Green function into Dyson Eq.(\ref{STEF})  
    leads to a transformed mass operator that is defined by 
      $\tilde M$ = $\left [ I - \frac {1}{3k^2_0}M \right ]^{-1}M$
    and gives possibility to get for the effective dielectric permittivity 
    an operator relation
\begin{equation}\label{LLF}
\left ( \varepsilon_{eff} + 2\varepsilon_0 \right )^{-1} 
\left ( \varepsilon_{eff} - \varepsilon_0 \right )
= - \frac {\tilde M}{3k^2_0}
\equiv \frac {4\pi}{3} \, \nu_{eff}
\end{equation}
which we call a generalized Lorentz--Lorenz formula. 
The transformed mass operator $\tilde M$ is evaluated by applying the Feynman diagrams technique 
\cite{FRISH66,FRISH67} or via equivalent method of asymptotic expansions \cite{FINK68,BAR76}.
      The mass operator asymptotic expansions are obtained from asymptotic expansions for
      ensemble averaged $\left\langle T \right\rangle$ electric field random scattering operator 
      $T$ of the medium and the equation 
      $\left\langle T \right\rangle$ = $\tilde M$ + $\tilde M \tilde G^{(0)} \left\langle T \right\rangle$.
According to definition \cite{NEW66}, the scattering operator  
      $T_{\alpha\beta} (\bf{r},\bf{r}')$
      satisfies the Lippmann--Schwinger type equation, 
      $T$ = $\tilde V$ + $\tilde V \tilde {G^{(0)}} T$,
      where the medium transformed scattering potential $\tilde V$ is equal to sum
      of single particle transformed potentials 
      $\tilde V_1(\bf{r})$ = -$k_0^{2}$ $3$ $\left[ \hat\varepsilon_1 (\bf{r}) - \varepsilon_0 \right]$ 
      / $\left[ \hat\varepsilon_1 (\bf{r}) + 2\varepsilon_0 \right]$,
      provided that the particles are not overlapped. 

\section{Generalized and classical Lorentz--Lorenz formula}
Making comparison between the obtained Lorentz--Lorenz formula (\ref{LLF}) and 
the classical one \cite{BORNW64} shows that a quantity $\nu_{eff}$ denoting the (\ref{LLF})
right hand side has sense of the medium effective electric susceptibility per unit volume 
tensor operator. 
   The ratio $\nu_{eff} / f_1$, with $f_1$ being the particle number density, is an electric 
   susceptibility per a scatterer. 
The left hand side of the generalized Lorentz--Lorenz formula is really looking like very 
familiar algebraic construction \cite{BORNW64} and in the right hand side of this one stands 
the transformed mass operator, which includes all significant items influencing the dielectric 
and conducting properties of composites. 
     Besides the frequency effects, i.e. accounting for intrinsic properties of constituency 
     by their complex dielectric functions, the mass operator includes geometric effects, 
     caused by resonant behavior of composite due to the shape and size of particles, 
     as well as cluster effects because of correlations between particles, which can result in 
     significant modification of particle scattering properties. 
The obtained Lorentz--Lorenz formula includes  such known cases as Maxwell--Garnett mixing rule 
with Clausius--Mossotti formula in the presence of multiple scattering on small correlated 
\cite{FINK64} and non--correlated \cite{MALET05} dielectric spheres. 
     The case of non--correlated dielectric particles is described by transformed mass operator 
     in the asymptotic form  $\tilde M$ $\approx$ $\int d1 f_1 T_1$ where $f_1$ is related to 
     a particle center $\bf {r_1}$ and $T_1$ is the single particle electric field scattering 
     operator with transformed scattering potential  $\tilde V_1 (\bf {r} - \bf {r_1})$.

\section{Effective magnetic permeability}
If statistical ensemble of particles is homogeneous and isotropic in average, 
the operators $\varepsilon_{eff}$  and $\tilde M$  are functions of spatial coordinates differences 
and can be expanded into Fourier integrals along spatial harmonics $\exp (i\bf {k} \bf {r})$. 
     Corresponding tensor Fourier transformations are defined by transversal $t$ and longitudinal 
     $\ell$  with respect to the wave vector $\bf {k}$ components, as
     $\varepsilon_{eff, \alpha \beta}(\bf {k})$ = $(\delta_{\alpha\beta} - \hat k_{\alpha} \hat k_{\beta})$
     $\varepsilon^{t}_{eff}(k)$ + $\hat k_{\alpha} \hat k_{\beta}$ $\varepsilon^{\ell}_{eff}(k)$,
     with $\hat k$ being the unit vector along $\bf {k}$. 
In terms of these components the Lorentz--Lorenz formula (\ref{LLF}) reads
\begin{equation}\label{LLFC}
\frac {\varepsilon^{t,\ell}_{eff}(k) - \varepsilon_0}{\varepsilon^{t,\ell}_{eff}(k) + 2\varepsilon_0}
= - \frac {\tilde M^{t,\ell}(k)}{3k^2_0}
\end{equation}
Actually, one does see in the left hand side of this formula the transversal and longitudinal 
effective dielectric permittivities \cite{LL04,AG06} for a monochromatic plane wave.

The spatial transversal and longitudinal Fourier components of the transformed mass operator 
in the right hand side of Eq.(\ref{LLFC}) take in the case of  non--correlated (independent) particles
the form $\tilde M^{t,\ell}(k)$ = $f_1$ $T^{t,\ell}(\bf {k},\bf {k})$. 
     Note, a quantity  $T_{\alpha\beta}(\bf {k},\bf {k'})$ denotes the electric field tensor 
     scattering operator of a single particle in spatial Fourier transform (momentum) representation 
     with arbitrary wave vectors $\bf {k}$ and $\bf {k'}$. 
After having been restricted on wave vector spherical surface  $\bf {k}$ = $\bf {k'}$ = $k_0$ , 
the double transversal part of this tensor with respect to directions of both scattering and incident 
wave vectors gives the tensor scattering amplitude in far wave zone of the particle \cite{NEW66}. 
     Nevertheless, we have met the tensor scattering operator $T_{\alpha\beta}(\bf {k},\bf {k})$ 
     for forward scattering direction, when the both wave vectors are identical and not equal to 
     the wave number in the host medium with respect to their absolute magnitudes. 
We made supposition that this value of the tensor scattering operator is defined by its transversal 
and longitudinal components $T^{t,\ell}(\bf {k},\bf {k})$  that is correct, for example,
in the case of a spherical particle.
   Following \cite{LL04} we denote
\begin{equation}\label{MUEF}
1 - \frac {1}{\mu_{eff}}
=
\frac {\omega^2}{c^2} \,
\frac {\varepsilon^{t}_{eff}\left ( k \right ) - \varepsilon^{\ell}_{eff}\left ( k \right )}{k^2}
=
\frac {\left [  \tilde M^t \left ( k \right ) - \tilde M^{\ell} \left ( k \right ) 
\right ]/k^2}{\left [ 1 + \tilde M^t \left ( k \right )/3k^2_0 \right ] \,  
\left [ 1 + \tilde M^{\ell} \left ( k \right )/3k^2_0 \right ]}
\end{equation}
setting next
$4\pi \bf{I}(\bf{k})$ = $(1 - 1/\mu_{eff})$$\left\langle \bf{B}(\bf{k}) \right\rangle$ 
and rewriting  the ensemble averaged second Maxwell's  Eq.(\ref{STM}) identically as 
\begin{equation}\label{STMI}
i\, \bf{k}\times 
\left (  \left \langle \bf{B}(\bf{k})\right \rangle - 4\pi \bf{I}(\bf{k}) \right )
= - \frac {i\omega}{c} \varepsilon^{\ell}_{eff}(\bf{k}) \left \langle \bf{E}(\bf{k})\right \rangle
\end{equation}
Eq.(\ref{MUEF}) defines in the limit $k$ $\rightarrow$ 0 an effective magnetic permeability $\mu_{eff}$, 
with  $\bf{I}(\bf{k})$ being a vector of effective medium magnetization. 
       In result, Eq.(\ref{STMI}) takes a form of the macroscopic Maxwell's equation 
       with magnetic flux vector \cite{JACK98}, given by
        $\left \langle \bf{H}(\bf{k})\right \rangle$ =  $\left \langle \bf{B}(\bf{k})\right \rangle$ 
        - $4\pi \bf{I}(\bf{k})$
       and dielectric  displacement current defined by longitudinal component of effective 
       dielectric permittivity.

\section{Electric dipole and magnetic dipole scattering}
Let us evaluate the effective magnetic permeability defined by (\ref{MUEF}) in the case 
of independent spherical particles. 
     Instead to find a solution to the Lippmann--Schwinger equation we will construct here   
     the single particle scattering operator $T_1$ in approximation of electric dipole 
     and magnetic dipole scattering. 
The incident electromagnetic wave induces into the particle the electric dipole and magnetic dipole moments
$\bf{p}$ = $\varepsilon_0 \eta \bf{E}_0$ and $\bf{m}$ = $\chi \bf{H}_0$ 
with dielectric and magnetic susceptibilities $\eta$ and $\chi$, respectively. 
     The Hertz's vector of electric dipole and magnetic dipole scattering
     $\Pi^{(el)}_\alpha (\bf{r})$ and $\Pi^{(mag)}_\alpha (\bf{r})$   
     is given according to \cite{STR41} by   
     $\Pi^{(el)}_\alpha (\bf{r})$ = - $(4\pi/\varepsilon_0)p_\alpha G_0 (r)$
     and 
     $\Pi^{(mag)}_\alpha (\bf{r})$ = - $(4\pi/ik_0\varepsilon^{1/2}_0)
     \text{e}_{\alpha\beta\gamma} m_\beta \nabla_\gamma G_0(r)$, 
     respectively, where $\text{e}_{\alpha\beta\gamma}$  is absolute anti-symmetric unit tensor. 
Bearing in mind the rule to evaluate the electric wave field via the Hertz' vector 
and also via the scattering operator,
$E$ = $G^{(0)} T_1 E^{(0)}$, one can get the following expressions for scattering operators 
of electric dipole and magnetic dipole scattering, respectively, in the form
\begin{eqnarray}
T^{(el)}_{\alpha\beta}(\bf{r},\bf{r}') = 
- \, 4\pi k^2_0 \, \eta \, \delta (\bf{r}) \delta (\bf{r}') \delta_{\alpha\beta},
\nonumber
\\
T^{(mag)}_{\alpha\beta}(\bf{r},\bf{r}') = 
- \, 4\pi \chi \, \text{e}_{\alpha\xi\eta} \, \text{e}_{\beta\xi\gamma} \, \nabla_\eta \nabla^{'}_\gamma 
\delta (\bf{r}) \delta (\bf{r}').
\label{OEMS}
\end{eqnarray}
The evident expressions of spherical particle electric and magnetic susceptibilities 
$\eta$ and $\chi$ are written out by comparison with Mie solution \cite{BORNW64} 
for electromagnetic wave scattering in all points of particle outside and have 
in Hulst \cite{HULST57} partial scattering amplitudes the form 
$\eta$ = $(3i/2k^3_0)a_1$ and $\chi$ = $(3i/2k^3_0)b_1$. 
The transversal and longitudinal components of the obtained scattering operators 
in momentum representation for forward scattering direction are given by
\begin{eqnarray}
T^{(el)t}(\bf{k},\bf{k}) = T^{(el)\ell}(\bf{k},\bf{k})
= - \, 4\pi \, \eta \, k^2_0 ;
\nonumber
\\
T^{(mag)t}(\bf{k},\bf{k}) = - \, 4\pi \chi \, k^2, \ 
T^{(mag)\ell}(\bf{k},\bf{k}) = 0.
\label{COEMS}
\end{eqnarray}
As one see, the electric dipole gives equal contributions into transversal and longitudinal components 
of the scattering operator in momentum representation for forward scattering direction 
but the magnetic dipole contributes into the transversal component only. 
     As consequence the difference,
     $\tilde M^{t}(k)$ - $\tilde M^{\ell}(k)$ = - $4 \pi f_1 \chi k^2$,
     includes contribution of magnetic dipole scattering only. 
As result the Eq.(\ref{MUEF}) for effective magnetic permeability takes a form
\begin{equation}\label{MUEFS}
1 - \frac {1}{\mu_{eff}} = 
\frac {4 \pi f_1 \chi}{\left [  1 - (4 \pi/3) f_1 \eta  \right ]^2}
\end{equation}

\section{Effective diamagnetism}
The electric and magnetic susceptibilities of spherical particle with radius $r_0$
small compared to the wavelength in the host medium are given according to \cite{BORNW64,HULST57}
by $\eta$ = $r^3_0$ and $\chi$ = - $(1/2) r^3_0$, respectively, in the limit of perfectly 
reflection because of high dielectric permittivity or conductivity of the particle. 
    These electric and magnetic susceptibilities of a spherical particle lead 
    to effective diamagnetism of random medium consisting of such particles due to the Eq.(\ref{MUEFS}). 
In the limit of small filling fraction, $f_1 r^3_0$ $<<$ 1, these equation gives formula
$\mu_{eff}$ $\approx$ 1 + $4\pi f_1 \chi$, obtained in \cite{SCHELK52} yet. 
     The Eq.(\ref{MUEFS}), with ignoring contribution of electric susceptibility in 
     the right hand side denominator, coincides in the form with a formula \cite{TAMM46} 
     for electron diamagnetism effect. 
The denominator in the right hand side of the Eq.(\ref{MUEFS}) leads to the diamagnetism 
enhancement because existence of effective dielectric permittivity given according to 
Eq.(\ref{LLFC}) by   
$(\varepsilon^{\ell}_{eff} - \varepsilon_0)$ / $(\varepsilon^{\ell}_{eff} + 2\varepsilon_0)$
= $(4\pi)/3)$ $f_1$ $\eta$.
One can say that diamagnetism is enhanced by appearance of additional effective dielectric 
displacement current in the medium.

\section{Conclusion}
In this paper the physical nature of diamagnetic property in the effective magnetic 
permeability at coherent electromagnetic wave multiple scattering by statistical 
ensemble of independent perfectly reflected non--magnetic small spherical particles 
has been revealed. 
   The diamagnetic physical nature consists in that the electric dipole scattering 
   by a particle gives equal contributions into transversal and longitudinal components 
   of the effective permittivity whereas the magnetic dipole scattering contributes 
   into the transversal component only. 
This observation is made with the aid of generalized Lerentz--Lorenz formula technique
for effective dielectric permittivity with spatial dispersion.

\section*{Acknowledgments} 
This work was supported in part by Russian Foundation for Basic Research, Grant 09-02-00920-a
and 09-02-12433-OFIM, by the Russian Academy of Sciences projects 
"Passive multichannel human radio- and  acousto-thermotomography  in near zone" and
"Investigations of new types of photonic crystals for the development of optoelectronic 
elements of infocommunicationic nets".

\end{document}